\documentclass[%
 reprint,
superscriptaddress,
 amsmath,amssymb,
 aps,
]{revtex4-2}

\usepackage{graphicx}
\usepackage{dcolumn}
\usepackage{bm}

\usepackage[normalem]{ulem}
\usepackage[pdftex,dvipsnames,usenames]{xcolor}
\usepackage{color,soul}

\begin{document}

\title{Experimental space-division multiplexed polarization-entanglement distribution through 12 paths of a multicore fiber}

\author{Evelyn A. Ortega}
	\email{evelyn.ortega@oeaw.ac.at}
	\affiliation{Institute for Quantum Optics and Quantum Information - IQOQI Vienna, Austrian Academy of Sciences, Boltzmanngasse 3, 1090 Vienna, Austria}
	\affiliation{Vienna Center for Quantum Science and Technology (VCQ), Vienna, Austria}
\author{Krishna Dovzhik}
	\affiliation{Institute for Quantum Optics and Quantum Information - IQOQI Vienna, Austrian Academy of Sciences, Boltzmanngasse 3, 1090 Vienna, Austria}
	\affiliation{Vienna Center for Quantum Science and Technology (VCQ), Vienna, Austria}
\author{Jorge Fuenzalida}
    \affiliation{Institute for Quantum Optics and Quantum Information - IQOQI Vienna, Austrian Academy of Sciences, Boltzmanngasse 3, 1090 Vienna, Austria}
	\affiliation{Vienna Center for Quantum Science and Technology (VCQ), Vienna, Austria}
	\affiliation{Current address: Fraunhofer Institute for Applied Optics and Precision Engineering IOF, Albert-Einstein-Str. 7, 07745 Jena, Germany}
\author{Sören Wengerowsky}
    \affiliation{Institute for Quantum Optics and Quantum Information - IQOQI Vienna, Austrian Academy of Sciences, Boltzmanngasse 3, 1090 Vienna, Austria}
	\affiliation{Vienna Center for Quantum Science and Technology (VCQ), Vienna, Austria}
	\affiliation{Current address: ICFO-Institut  de  Ciencies  Fotoniques,  The  Barcelona  Institute  of Science  and  Technology,  08860  Castelldefels  (Barcelona),  Spain}
\author{Juan Carlos Alvarado-Zacarias}
    \affiliation{CREOL, The University of Central Florida, Orlando, Florida 32816, USA}
\author{Rodrigo F. Shiozaki}
    \affiliation{Departamento de F\'isica, Universidade Federal de S\~{a}o Carlos, Rodovia Washington Lu\'is, km 235—SP-310, 13565-905 S\~{a}o Carlos, SP, Brazil}
\author{Rodrigo Amezcua-Correa}
    \affiliation{CREOL, The University of Central Florida, Orlando, Florida 32816, USA}
\author{Martin Bohmann}
    \email{martin.bohmann@oeaw.ac.at}
    \affiliation{Institute for Quantum Optics and Quantum Information - IQOQI Vienna, Austrian Academy of Sciences, Boltzmanngasse 3, 1090 Vienna, Austria}
	\affiliation{Vienna Center for Quantum Science and Technology (VCQ), Vienna, Austria}
\author{Rupert Ursin}
    \email{rupert.ursin@oeaw.ac.at}
	\affiliation{Institute for Quantum Optics and Quantum Information - IQOQI Vienna, Austrian Academy of Sciences, Boltzmanngasse 3, 1090 Vienna, Austria}
	\affiliation{Vienna Center for Quantum Science and Technology (VCQ), Vienna, Austria}

\begin{abstract}
The development and wide application of quantum technologies highly depend on the capacity of the communication channels distributing entanglement.
Space-division multiplexing (SDM) enhanced channel capacities in classical telecommunication and bears the potential to transfer the idea to quantum communication using current infrastructure.
Here, we demonstrate an SDM of polarization-entangled photons over a $411\,$m long 19-core multicore fiber distributing polarization-entangled photon pairs through up to 12 channels simultaneously.
The quality of the multiplexed transfer is evidenced by high polarization visibility and CHSH Bell inequality violation for each pair of opposite cores.
Our distribution scheme shows high stability over 24 hours without any active polarization stabilization and can be effortlessly adapted to a higher number of channels.
This technique increases the quantum-channel capacity and allows the reliable implementation of quantum networks of multiple users based on a single entangled-photon pair source.
\end{abstract}

\date{\today}
\maketitle

\section{Introduction}
Quantum communication has been rapidly progressing in private and secure information transmission \cite{wootters1982nocloning, ekert1991,gisin2002Quantumcryptography,xu2020}.
In any quantum communication scheme, one needs communication channels to distribute the quantum information encoded in properties of photons between the users.
The major types of channels are free-space links \cite{ursin2007,yin2012}, satellite \cite{liao2018satellite} and optical fibers \cite{korzh2015,boaron2018secure,wengerowsky2020passively},
where fiber connections are the easiest and already deployed option for the last mile connection to the customers.
In this context, a vital aspect of optimizing resources is the development and implementation of high-capacity quantum channels.

Today multiplexing is widely used in classical telecom infrastructure \cite{wang2015experimental} transmitting data in parallel to increase the capacity of a single deployed channel.
Space-division multiplexing (SDM) is currently being investigated in the classical communications context \cite{richardson2013}.
This multiplexing approach relies on few-mode \cite{sillard2014few,van2018138}, multimode \cite{sillard201650} and multicore fibers \cite{van2014ultra,ryf2016long,Saitoh2016}, as well as novel mutiplexers and demultiplexers \cite{velazquez2015six,alvarado2018mode} to harness the spatial dimension of light and boost capacity compared to single mode fiber systems.
In quantum communication schemes, multiplexing technologies have been implemented in different degrees of freedom, such as time \cite{Chen_2009}, wavelength \cite{wengerowsky2018}, and polarization \cite{chen2017}.
However, the multiplexing of photons in terms of their spatial modes over optical fibers has not been thoroughly explored yet in an experiment.
In terms of fibers links, single-mode fibers (SMF) are not suitable for SDM implementations since only one spatial mode can be carried.
In contrast, the fibers capable of transmitting multiple spatial modes, such as a multi- and few-mode fibers, require a complex system \cite{leedumrongwatthanakun2020programmable} to avoid the mode-coupling effects \cite{cui2017} and are susceptible to external influences \cite{loeffler2011} inhibiting long-term stability.
A multicore fiber (MCF) is superior in this sense because all the spatial modes are guided through individual cores that share the same cladding \cite{Tang2019}.
The MCF preserves the distinguishability of discrete propagation paths while assuring the low cross-talk between the cores.

In the context of quantum information, MCF recently gained popularity \cite{xavier2020quantum}, where, e.g., the parallel transmission of quantum and classical signals has been shown \cite{Dynes:16, Bacco2019}.
However, improving quantum-communication applications, it is necessary to device strategies which enable the parallel transmission of quantum information over the same channel.
As a consequence of the multiple paths available in MCF, they have been a useful tool for exploring high-dimensional quantum communication \cite{canas2017high,ding2017high,DaLio2021,Hu:20}, transmission of four-dimensional spatially-entangled photon pairs \cite{Lee2017,Lee:19}, and generation of multidimensional entanglement state \cite{PhysRevApplied.15.034024}.
However, the huge potential of exploiting quantum correlations in different degrees of freedom for MCF communication has not been tackled so far.

Here, we present an experimental implementation where a MCF is used as a communication channel to simultaneously distribute multiple entangled photon pairs independently of each other from a single entangled-photon pair source.
We report on the characterization of the coupling and distribution of polarization-entangled photons through a 19-core fiber (19-MCF), demonstrating excellent performance of the system.
In particular, we certify high path visibilities, i.e., the quality of correlations between opposite fiber cores, and that the polarization entanglement between two opposite cores is well preserved over the MCF as well as in the fiber-splice-based separation of the paths.
We demonstrate long-term stability, relevant for applications, of the quantum SDM setup.
To the best of our knowledge, this is the first experimental realization of a polarization-entanglement distribution scheme based on momentum correlation over MCF and the simultaneous use quantum correlations in multiple degrees of freedom.
Our results show that quantum SDM can be achieved by exploring intrinsic momentum correlations in the spontaneous parametric down-conversion (SPDC) process, which is not possible via random distribution of the photons.

\section{Implementation}
\begin{figure} [h]
   \includegraphics[width=1\columnwidth]{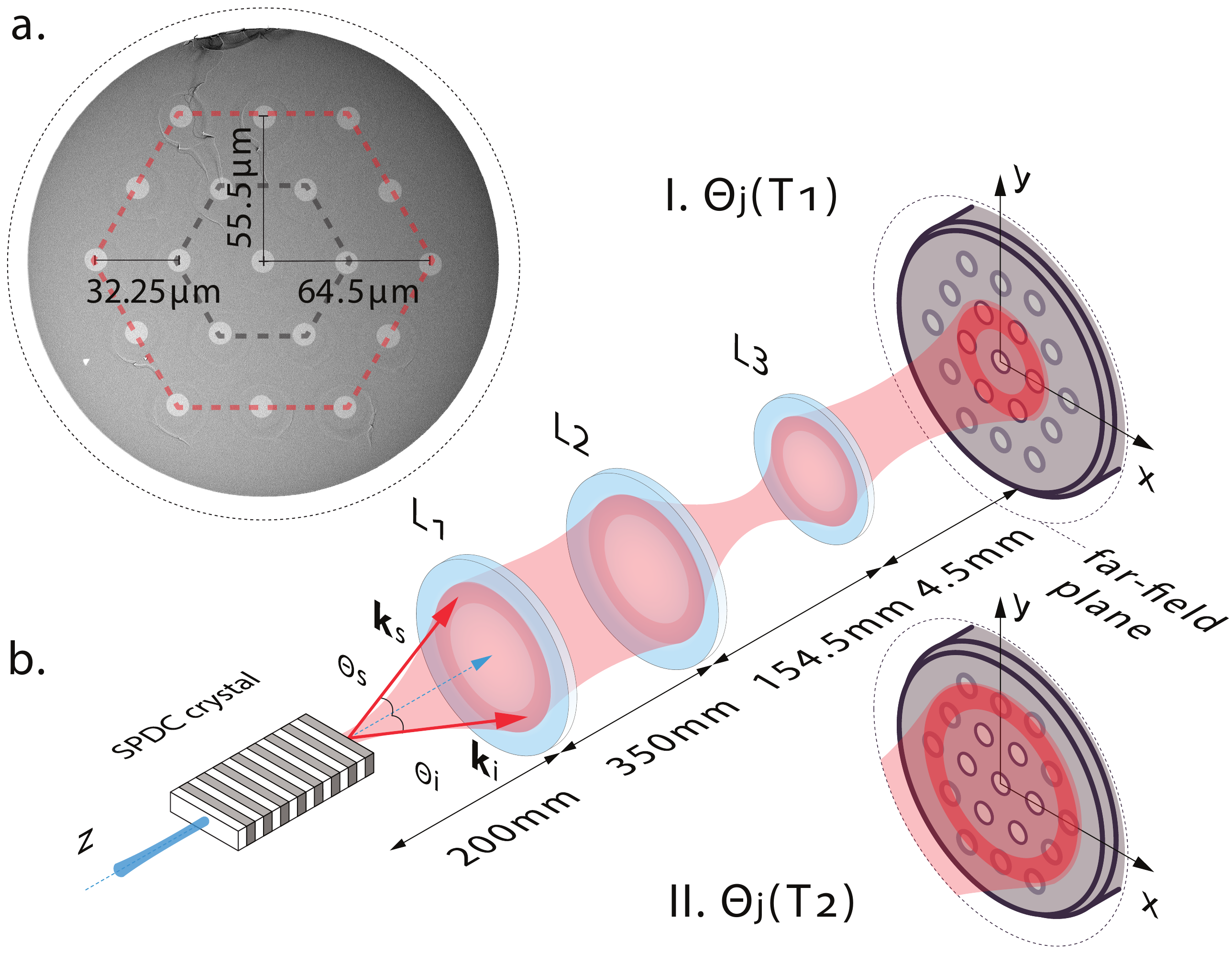}
   \caption{Illustration of the implemented SDM through a 19-MCF. (a) An electron microscope image of the 19-MCF.
   The inner and the outer ring of cores are highlighted by the black and red hexagons respectively.
   (b) A laser beam (blue, $z$ is the pumping direction) pumps a non-linear crystal where the down-converted photons are created.
   The emitted photons form an emission cone (non-collinear type-0 phase matching) shown in red.
   Opening of the cone defined by the angles $\Theta_s$ and $\Theta_i$ for the signal and idler photons depends on the crystal temperature~$\text T$. 
   The lenses $\text L_{1}$, $\text L_{2}$ and $\text L_{3}$ image and demagnify the far-field plane of the crystal onto the MCF's end-face.
   We show two cases for different cone openings $\Theta_j(\text T)$ for the temperature: (I.) $\text T_1$ illuminating the inner ring of cores; (II.) $\text T_2$ illuminating the outer ring of cores. Fiber end-faces are shown not to scale.}
   \label{fig:mcf-spdc}
    \end{figure}

The experimental setup is schematically depicted in Fig. \ref{fig:mcf-spdc} b).
Photon pairs were generated by a type-0 phase-matched SPDC source in a Sagnac configuration (see Fig. \ref{fig:setup} in Appendix \ref{appendix:source} for more details).
Through the non-linear crystal operated in the Sagnac configuration, a polarization-entangled state is created.
The momentum conservation in the SPDC process warrants that the signal and idler photons are emitted with opposite transverse components of the wave vectors ($\mathbf{k}_s$ and $\mathbf{k}_i$) at the telecommunication wavelength.
Hence, the photons of an entangled pair are located at diametrically opposite points in the far-field plane of the crystal, i.e., the transverse momentum representation of the photon pair \cite{PhysRevA.75.042317, PhysRevLett.103.160401}.
By a cleverly devised lens system (see Fig. \ref{fig:mcf-spdc} b), we imaged the far-field plane of the SPDC crystal to the end-face of the MCF.
Thus, opposite cores of the MCF were linked via the momentum correlations exhibited by the photon pairs.
The resulting quantum state inside the MCF is a polarization-spatial hyperentangled state
\begin{align}
    |\boldsymbol{\Phi}\rangle=|\Phi^+\rangle_{\mathrm{pol}}\otimes \sum_{m=1}^N g_m|c_m,c_m^{\prime}\rangle_{\mathrm{cores}},
\end{align}
where the first part is the polarization-entangled Bell state and the latter one describes the spatial-entangled state as a high-dimensional superposition of opposite cores (paths), $c_m$ and $c_m^{\prime}$.
Here, $g_m$ are normalized probability amplitudes, $\sum_{m=1}^N |g_m|^2=1 $, and $N$ is the number of opposite cores.

The lens $\text{L}_1$ performed a Fourier transform of the transverse position of the photons at the crystal plane (see. Fig. \ref{fig:mcf-spdc} b), therefore, the far-field was placed at the focal point of the lens $\text{L}_1$.
The far-field plane was imaged and demagnified to match it to the end-face of the MCF using a $2f/2f'$-imaging system consisting of the lenses $\text{L}_2$ and $\text{L}_3$ (see Appendix \ref{appendix:source} for more details).
The precise matching of the SPDC emission cone onto the rings of the MCF was achieved through temperature control of the non-linear crystal.
Here, we take advantage of the fact that the temperature influences the phase-matching condition of the SPDC process resulting in a varying opening angle of the emission cone \cite{Lee15, ortega2020} ($\mathbf{k}_s$ and $\mathbf{k}_i$ in the Fig. \ref{fig:mcf-spdc}, b).

The 19-MCF was fabricated at CREOL in the University of Central Florida, where stack and draw methods were used in the fabrication process.
A set of 19 trenches-assisted core rods were placed in a hexagonal array inside a jacketing tube, then a set of packing rods of different sizes made out of silica were used to fill up the space between the core rods.
Once the preform is assembled, the fiber was fabricated in our drawing tower.
The fiber was designed to support the fundamental mode at $1550\,$nm, its cores have a diameter of $\sim 8.5\,\mu$m with an index difference $\Delta$n=$5.5\times10^{-3}$ with respect to silica, whereas the trench has an index difference $\Delta$n=$-5\times10^{-3}$ with respect to silica.
The 19 cores of the $411\,$m long MCF are arranged in an array with hexagonal symmetry with a core-to-core distance of $\sim 32.25$ $\mu$m (see Fig. \ref{fig:mcf-spdc}, a).
The array of 19 cores naturally consists of a central core and two hexagonal rings of cores to which we refer to as the inner ring (6 cores closest to the central core) and the outer ring (12 cores at the circumference of the fiber end-face) (see Fig. \ref{fig:mcf-spdc}, a).
A trench structure was implemented to reduce inter-channel cross-talk.
The MCF was connected to a fan-in/fan-out (FIFO) device which interfaced each core to a SMF with an insertion loss  $\leq3$ dB per core.
The FIFO was fabricated following the process described in \cite{MCF-JC}.
Through a careful alignment of the MCF with respect to the emission cone of the SPDC source (see Appendix \ref{appendix:source}) we realized a symmetrical intensity distribution at its end-face and we can reliably change the phase-matching to emit a cone to either illuminate the inner or the outer ring by changing the temperature of the crystal.

To characterize the correlation between the distributed entangled photons in the spatial and polarization degrees of freedom, we used two detection configurations where the single channels of the MCF were either connected to photon detectors directly or pair-wise through a pair of polarization analyzing modules, respectively (see Appendix \ref{appendix:source} for details).
Superconducting nanowire single-photon detectors (SNSPD) with an efficiency $\sim80\%$ and the dark-count rates of $\sim10^2$ Hz were used for the photon detection.
The detection events had been recorded by a time-tagging module for further analysis.
Correlated photon pairs were identified through coincidences in counts with the help of the temporal cross-correlation functions.

\section{Path visibility}

We start with the characterization of spatial correlations of the photons distributed over the MCF.
For this purpose, we measured the photon intensities and the coincidence counts between all possible combinations of cores in each ring (Fig. \ref{fig:setup}, module \textbf{a}).
All combinations of core pairs, 15 pairs for the inner ring and 66 pairs for the outer ring, were measured for $30\,$s each.

The SPDC source produces spatially entangled photons which show anti-correlations in momentum \cite{walborn2010spatial}.
The transverse wave vectors of the signal and idler photons are mapped through the far-field measurement onto MCF cores (paths) (see also Fig. \ref{fig:mcf-spdc}, b).
In this way, we relate the momentum correlations to the path correlations between opposite cores of the rings.

 \begin{figure}[h]
 \includegraphics[width=1\columnwidth]{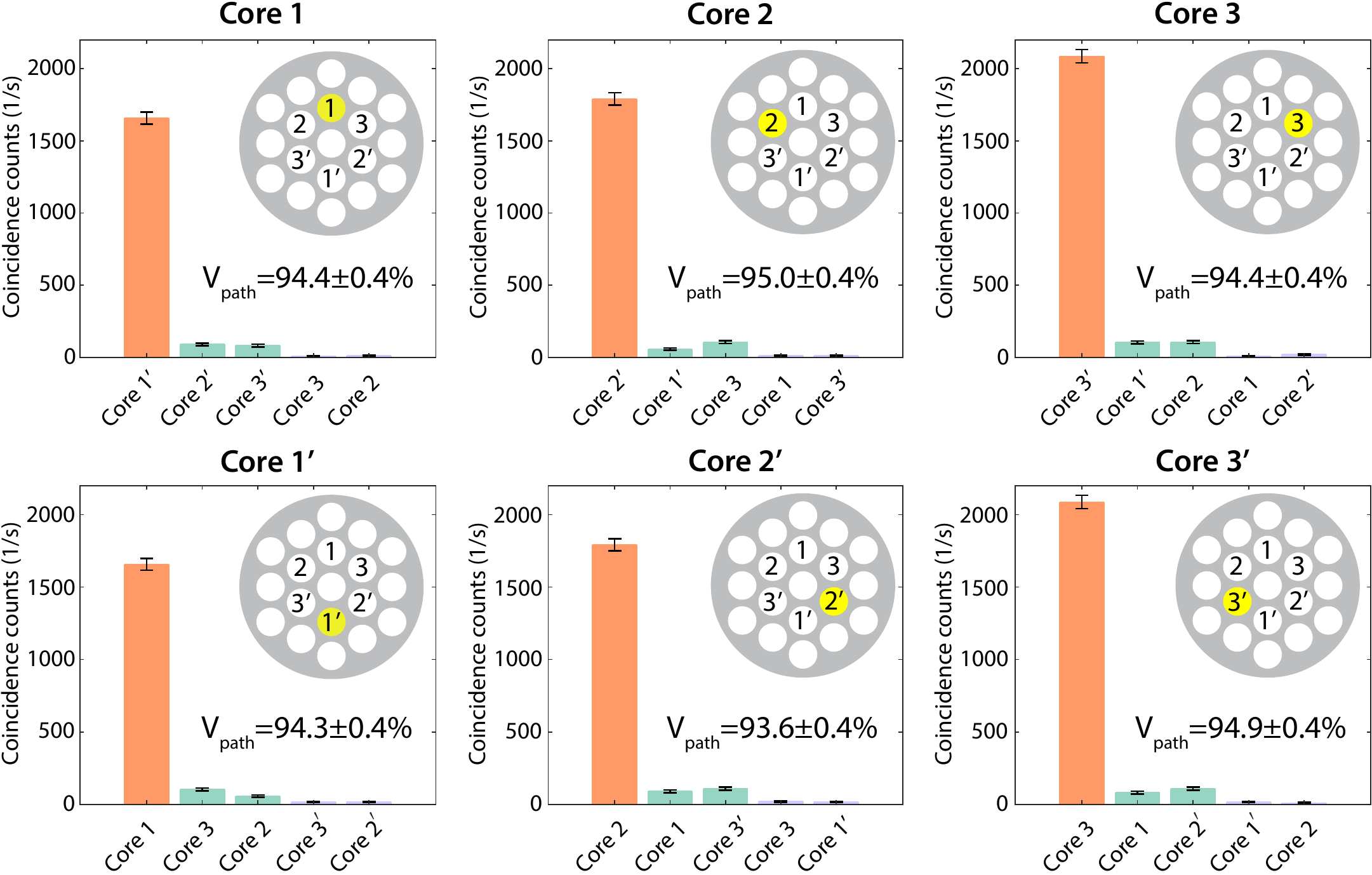}
 \caption{Coincidence-based correlation measurements for the cores of the inner ring of the 19-MCF.
 Each graph shows the coincidence counts measured between the core $m$ (highlighted in yellow) and all the other cores in the inner ring.
 The orange bar indicates the maximum coincidence counts observed between the opposite cores $m$ and $m'$.
 The green bars are the coincidence counts between the cores $m$ and the neighbor cores of the core $m'$.
 The last two bars in purple correspond to the coincidence counts between core $m$ and its neighbor cores.
 Pronounced correlations between opposite cores are reflected by the obtained values of the path visibilities indicated in each graph. 
 The experimental error bars were calculated assuming Poissonian counting statistics.}
 \label{fig:hist_innerring}
 \end{figure}
 
\begin{figure}[t]
\includegraphics[width=1\columnwidth]{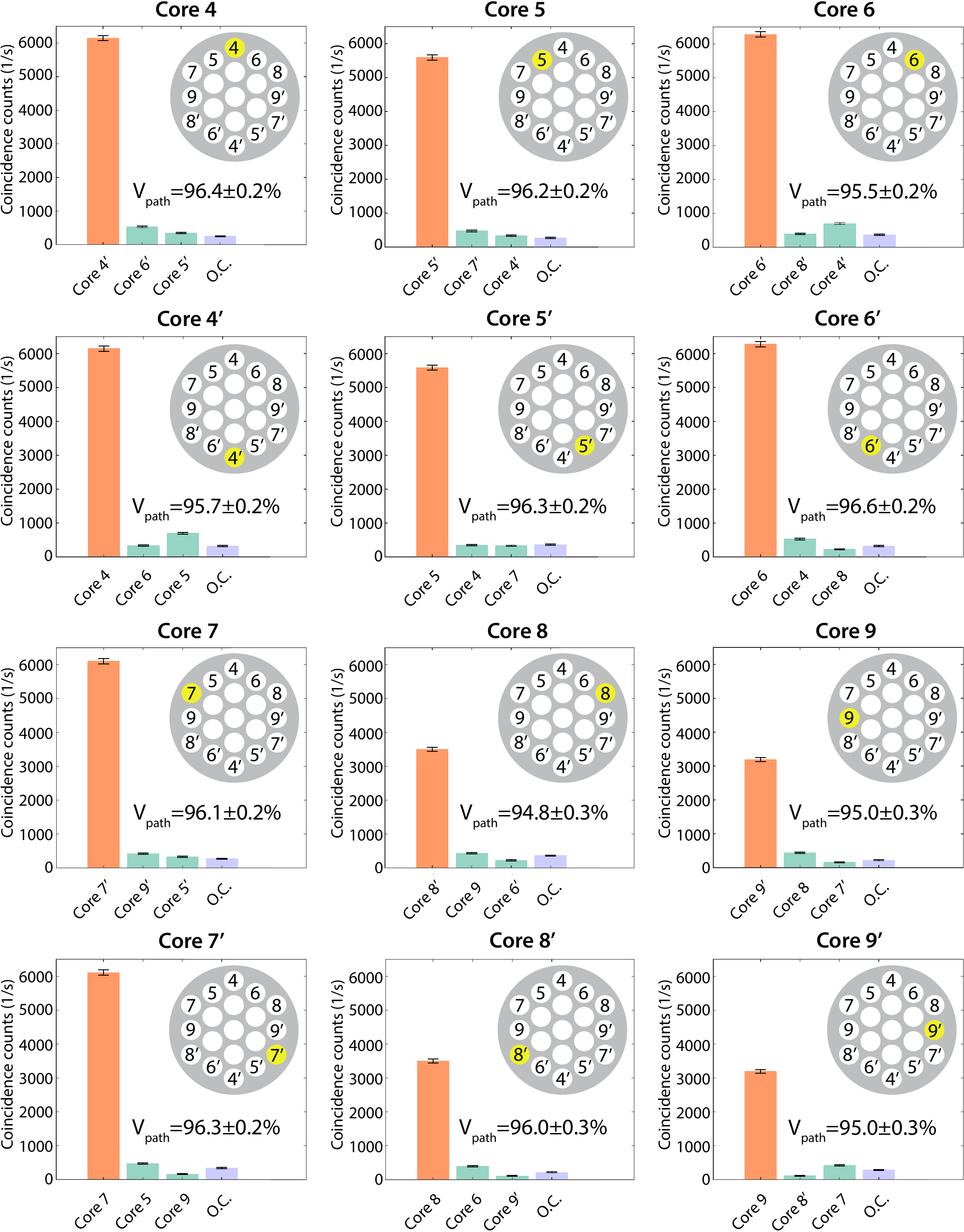}
\caption{Coincidence-based correlation measurements for the cores of the outer ring of the 19-MCF.
Each graph shows the coincidence counts measured between the core $m$ (highlighted in yellow) and all the other cores in the outer ring.
The orange bar indicates the maximum coincidence counts observed between the opposite cores $m$ and $m'$.
The green bars are the coincidences between the cores $m$ and neighbor cores of the core $m'$.
The last purple bar is the sum of the other combinations (O.C.) coincidence counts between the core $m$ and its other 8 neighbor cores in the outer ring.
This indicates strong correlations between opposite cores.
The path visibility of each core is indicated in the graphs.
The experimental error bars were calculated by assuming Poissonian counting statistics.}
\label{fig:hist_outerring}
\end{figure}

The coincidence measurements between the different cores are shown in Fig. \ref{fig:hist_innerring} (inner ring) and Fig. \ref{fig:hist_outerring} (outer ring).
Indeed, we observe the expected strong correlations between the opposite cores $m$ and $m'$ while almost no coincidences were detected between any non-opposite cores.
Although the MCF with the fan-out device shows different coupling efficiencies and transmission losses for each core, the behavior is equivalent for each core: high-correlations with the opposite core and low cross-correlations with the remaining cores.

To quantify these results, we introduce a measure which evaluates these correlations and the amount of cross-talk.
In analogy to the polarization visibility \cite{wolf2007introduction}, we introduce a corresponding measure, \textit{path visibility}, which is defined by the coincidence counts between all possible combinations of cores within the considered ring.
\begin{equation} 
\label{eq:path_vis}
        V_{\mathrm{path}}(m) =
         \frac{C_{mm'}-\frac{\sum_{l\neq m,m'}C_{ml}}{N}}{C_{mm'}+ \frac{\sum_{l\neq m,m'}C_{ml}}{N}}
    \end{equation}
where $V_\mathrm{path}(m)$ is the path visibility associated with the core $m$, $C_{mm'}$ is the number of coincidences between the core $m$ and the opposite core $m'$, $C_{ml}$ ($l\neq m, m'$) is the number of coincidences between the core $m$ and the other cores in the same ring.
Here the weighting of the sum by $1/N$, $N$ being the number of non-opposite combinations, assures a balanced consideration of wanted (opposite) and unwanted (non-opposite) correlations.
A high value of $V_\mathrm{path}(m)$ close to $1$ demonstrates strong correlations between opposite cores ($m$ and $m'$) and low cross-talk with the other cores.

The high path correlations in our experiment are reflected by the obtained values of the path visibilities.
Our measurements show that the path visibility was greater than $(93.6\pm0.4)\%$ for all the inner cores (1-3') and greater than $(94.8\pm0.3)\%$ for all the outer cores (4-9'), where we reach values up to $(96.6\pm0.2)\%$.
We have thus confirmed that the photon pairs are coupled to and distributed over the MCF in a way that the spatial correlations in the SPDC emission are mapped to the cores (path).
Importantly, this enables multiphoton experiments with up to twelve photons simultaneously using the outer ring of our MCF. 
\section{Polarization visibility}

Once we verified that our setup features the reliable distribution of spatially-correlated photon pairs through the MCF, we measured the quality of the polarization entangled state between opposite cores (see equation \eqref{eq:state} in Appendix \ref{appendix:source}).
For this purpose, we carried out a series of polarization visibility \cite{wolf2007introduction} measurements on opposite channels in horizontal/vertical (H/V) and diagonal/anti-diagonal (D/A) bases.
For this measurement, polarization analyzing modules were connected to opposite cores (Fig. \ref{fig:setup}, module \textbf{b}).
The polarization modules consist of a half-wave plate (HWP) and a polarizing beam-splitter (PBS) with a SNSPD connected to the transmitted and reflected output ports of the PBS.
Through polarization compensation, two mutually unbiased bases are set and no active control or compensation is needed. 
For each pair of cores and analyzer setting, 3 pairs for the inner ring and 6 pairs for the outer ring, we accumulated the data for $60\,$s.

\begin{figure}[h]
\includegraphics[width=1\columnwidth]{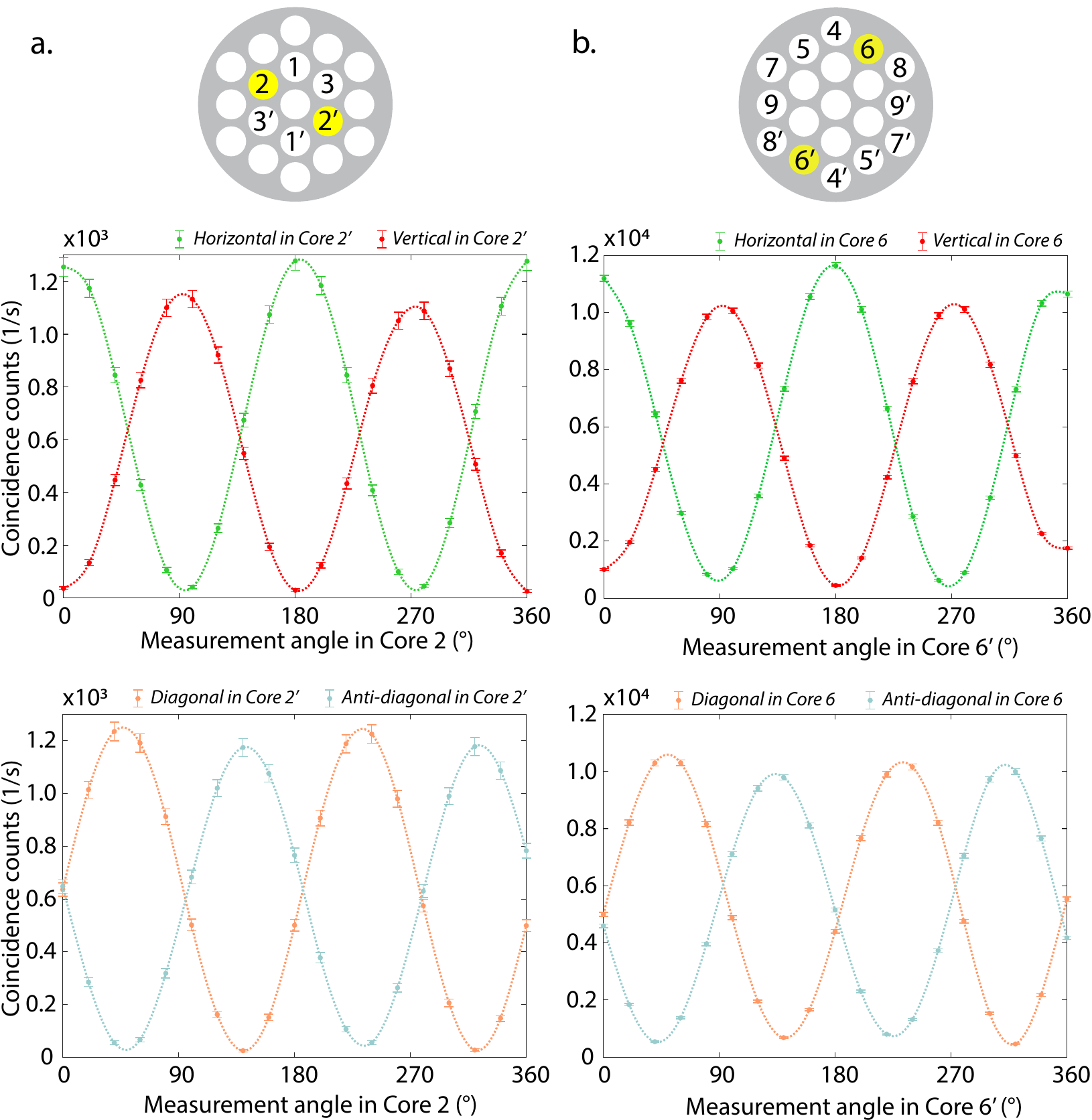}
 \caption{Coincidence count rates of polarization-entangled photon pairs.
 For each opposite cores 2-2' (inner ring, a.) and 6-6' (outer ring, b.), the coincidence counts were recorded for four different measurement angles [H (green), V (red), D (orange) and A (blue)] set in one polarization module (core 2' and core 6) while step-wise changing the angle in the second polarization module (core 2 and core 6'). 
 The error bars were calculated by assuming Poissonian counting statistics.}
 \label{fig:polvis}
 \end{figure}

To study the dependence on the polarization-measurement settings, one polarization module was set in H or D, while the second module had been scanned for the full range of the HWP angle, from $0^{\circ}$ to $360^{\circ}$ in steps of $20^{\circ}$.
The recorded coincidence data together with the corresponding fits are shown in Fig. \ref{fig:polvis}, for the inner and the outer ring.
The expected oscillatory behavior for polarization entangled states is observed.

In order to certify the polarization entanglement, we measured the polarization visibility for each pair of opposite channels (see Appendix \ref{appendix:allignment} for details).
All the combinations exhibit visibilities greater than $(84.20\pm0.04)\%$ in the H/V bases and $(83.60\pm0.04)\%$ in the D/A bases.
We have reached up to $(95.15\pm0.04)\%$ in the best case.
In all cases, the polarization visibilities are above $81 \%$, the threshold of quantum key distribution (QKD).
Minor differences in visibilities are caused by the fact that the cores and the coupling into the cores of the MCF are not perfectly identical \cite{MCF-JC}.

\section{Further analysis and applications}

In this section, we provide further analysis on the transferred entangled states and their applications.
Firstly, we calculate lower bounds on the quantum-state fidelities with respect to a maximally entangled state.
Secondly, we analysis the achievable secure key rate of our setup in a QKD implementation.
Thirdly, we perform a Bell-type experiment violating a CHSH inequality.
Furthermore, we provide a long-term stability analysis of the setup.

\subsection{Quantum-state fidelity}

In order to provide further analysis of the created and transmitted entangled states, we consider the quantum-state fidelity with the maximally entangled target Bell state $|\Phi^+\rangle_{\mathrm{pol}}$.
From the measurements of non-classical polarization-correlation visibilities in the HV and DA basis for each pair of opposite channels, we can calculate a lower bound on the Bell-state fidelity given by $\mathcal{F}_{\mathrm{lower}}=(V_{HV}+V_{DA})/2$, see, e.g., \cite{Blinov2004,Chang2016}.
This allows us to provide a (conservative) evaluation of the distributed entangled states.
The minimal obtained value of the lower bound on the fidelity was $(83.90\pm0.04)\%$.
In the inner ring the maximal value of $\mathcal{F}_{\mathrm{lower}}$ is $(94.56\pm0.04)\%$ and in the outer ring we reach a value of $(94.64\pm0.02)\%$.
These obtained lower bounds on the fidelity, further underline the quality of the transferred polarization entanglement.

\subsection{Achievable secure key rates}

So far, we have focused on the quality of the distributed quantum correlations through the MCF.
Now, we will determine the achievable secure key rate if our setup is used in a QKD scenario.
Here, it is important to stress that it is possible to create a key between each pair of opposite cores.
Thus, it is possible to implement a multiplexed communication system, enabling a high-throughput entanglement-based QKD.
The total secure key rate is, therefore, given by 
\begin{align}
    R^{\mathrm{s}}_{\mathrm{tot}}=\sum_m^{N} R^{\mathrm{s}}_{m},
\end{align}
where $R^{\mathrm{s}}_{m}$ is the secure key rate achievable between the $m$th pair of cores and $N$ is the total number of opposite cores.
We calculate each $R^{\mathrm{s}}_{m}$ based on the recorded coincidences and polarization visibilities (determining the quantum bit errors); cf. Appendix \ref{appendix:SKR} for details.
From the inner and outer ring, we achieve an averaged key rate per pair of $2.3\,$kHz and $5.7\,$kHz, respectively.
By exploiting our SDM scheme, we therefore obtain total key rates $R^{\mathrm{s}}_{\mathrm{tot}}$ from the inner and outer ring of $7$ kHz and $34$ kHz, respectively.
The difference in the key rates of the inner and outer ring can be understood by the different crystal temperatures of both settings, which SPDC emission and spatial correlation.
Our results clearly show that the achievable key rate scales with the number of used pairs of cores, demonstrating the advantage and potential of quantum-state multiplexed QKD through MCF.

\subsection{Bell inequality violation}
We further measured the Bell-type CHSH inequality in each pair of channels (see Appendix \ref{appendix:CHSH} for more details).
The polarization modules were prepared in the theoretical optimum values \cite{clauser1969proposed}, whose angles are ($22.5^{\circ}$, $67.5^{\circ}$) and ($0^{\circ}$, $45^{\circ}$).
With the combination of the four bases in both modules, we measured the CHSH inequality, which was violated in both rings.
In every pair of opposite cores, the S-parameters obtained were higher than $2.30\pm0.01$, which is an evidence that the quantum features are preserved through the MCF.
Moreover, we have reached a maximum value of $2.73\pm0.03$, which corresponds to the pair of cores with the highest visibility, $(93.96\pm0.04)\%$ in the H/V bases and $(95.15\pm0.04)\%$ in the D/A bases.

\subsection{Long-term stability}
\begin{figure}[h!]
    \includegraphics[width=1\columnwidth]{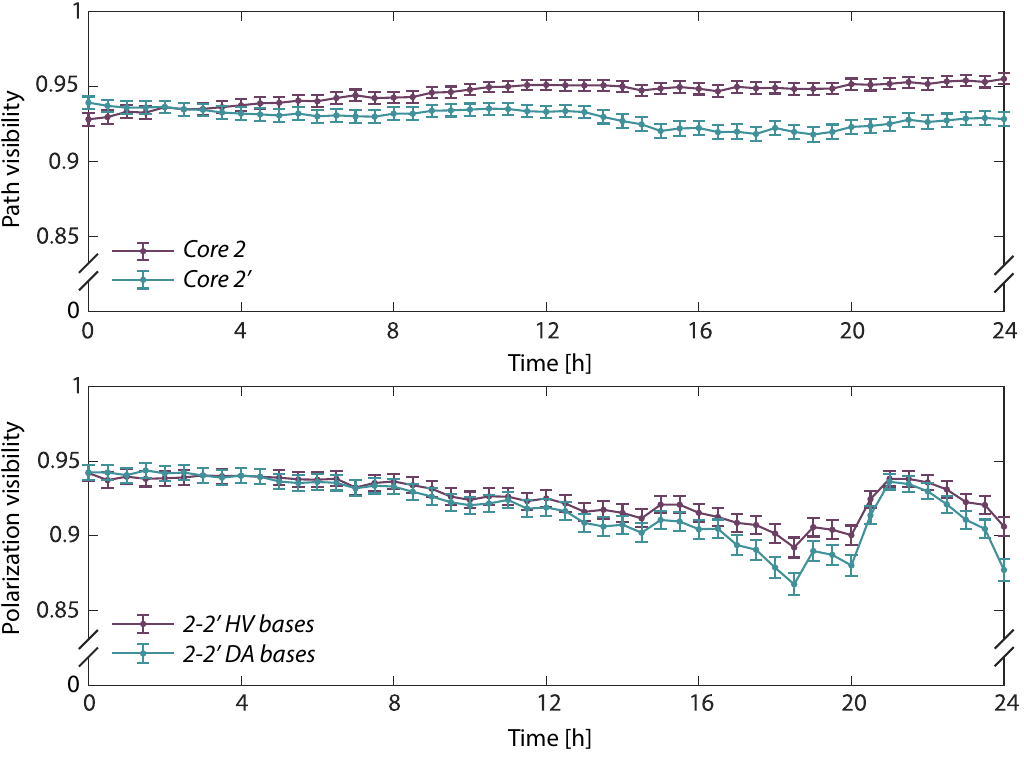}
    \caption{Long-term measurements on the entangled photon state over $411\,$m long MCF.
    A measurement for the path (upper plot) and polarization (lower plot) visibilities over a period of 24 hours for one pair of the inner ring core.
    The visibilities were calculated from the measurement in each bases integrated over 60 s, and the error bars were calculated by assuming Poissonian counting statistics for each point.
    The results demonstrate a stable performance of our MCF implementation for the distribution of correlated photon pairs.}
    \label{fig:24hrsvisibilities}
\end{figure}

In order to demonstrate the stability of our setup, we performed two long-term measurements of the path and polarization visibilities for one pair of cores over a period of 24 hours.
During this period, the coincidence counts were collected every $30\,$min for a period of $60\,$s.
For the polarization measurement, the visibilities in the H-V and D-A bases were measured alternately. 
As shown in the Fig. \ref{fig:24hrsvisibilities}, both polarization and path visibilities were stable with only marginal fluctuations for at least 24 hours without any external alignment or active stabilization. 
In the case of the path visibility, the minimum value was $(91.8\pm0.5)\%$ and the maximum $(95.5\pm0.4)\%$. 
The polarization visibility stayed in the range of values from $(86.8\pm0.7)\%$ to $(94.4\pm0.5)\%$.
The residual changes in the laboratory environment do change the polarization state more severely as the path visibility.
\newpage
\section{Discussion}
We have successfully implemented the distribution of polarization entanglement multiplexed in the path degree of freedom through a 19-core MCF.
We experimentally demonstrated the high quality with which the spatial modes are independently coupled into the MCF and quantified the resulting spatial correlations by high values of the introduced measure of path visibility of up to $(96.6\pm0.2)\%$.
Furthermore, we certified polarization entanglement between each opposite channel through high polarization visibilities and violating the CHSH inequality.
That visibility was maintained over a spooled $411\,$m long MCF without any active polarization compensation.
We illustrated the stability of our setup given by path and polarization visibility measurements over 24 hours, which is an important characteristic for future applications in real-world conditions outside the laboratory.

Exploiting the existing transverse momentum correlations in the emission of the SPDC source, we were able to collect polarization-entangled photon pairs into several pairs of opposite channels from a single SPDC source.
Hence, our scheme is based on the parallel employment of quantum correlations in several degrees of freedom.
The designed lens configuration allows us to access spatially separated rings and avoids cross-talk between neighboring cores by adapting the proper crystal temperature. 
Importantly, the achieved polarization visibilities allow the implementation of quantum information applications such as cryptographic protocols.
Our scheme provides a substantial advantage in terms of resources and versatility.
Due to the scalability and ease of upgrading without changing the entanglement source or the type of quantum state produced.
Therefore, the presented experiment opens an efficient pathway to SDM in quantum information protocols.

An essential requirement for the multiplexed transfer of polarization-entangled photon pairs through the MCF is that the entangled photon pairs can be identified as a pair at the receiver.
We realized this by mapping the momentum correlations of the photons to the cores of the MCF.
Notably, this cannot be achieved by just randomly distributing the SPDC photons (without exploiting additional quantum correlations) through the MCF.

Our experiment can be efficiently adapted to other types of multicore fibers \cite{Tang2019} and extended to using more cores at the same time by optimizing the source characteristic for strong momentum correlation, i. g., by changing the crystal length or pump waist.
As well as the entanglement source can be adapted with different crystal configurations to generate photon pairs in more than one emission cone \cite{PhysRevA.79.030301}.
Additionally, the development of new MCF's with a higher number of cores allows for an increase in the number of independently distributed photon pairs \cite{7936932}.
At the same time, the phase-matching conditions remain a tunable parameter allowing for the reliable collection of photons into more channels.

Quantum SDM in optical fibers opens access to implementing QKD schemes for high throughput quantum communications.
Besides this, our implementation of the SDM with polarization-entangled photons can be combined with other degrees of freedom, e.g., wavelength \cite{Pseiner_2021} or time-bin \cite{cui2017}, for enhancing communication capacity, and can be adapted for multi-user quantum network applications \cite{wengerowsky2018, Joshi2020}.
Furthermore, it provides a feasible way of implementing QKD based on position-momentum variables without time-consuming fiber scans \cite{almeida2005} as all modes are guided simultaneously through the fiber.
The next step is to ensure compatibility of the implemented SDM optical network with quantum technological applications, such as quantum communication systems.

Moreover, the presented experiment enables a variety of experimental applications:
in combination with MCF-integrated multiport beam-splitters \cite{Carine:20}, our experiment can be extended to a platform for quantum network applications.
And the simultaneous coherent collection of several photons could be used for multiphoton interference experiments \cite{PhysRevLett.118.153602,menssen2017} or the implementation of multimode multi-partner quantum metrology \cite{albarelli2020}.

\section*{Acknowledgments}
We acknowledge funding by the European Union’s Horizon 2020 programme grant agreement No.857156 (OpenQKD) and the Austrian Academy of Sciences.
J.C.A.-Z. and R.A.-C. acknowledge support US. Army Research Office W911NF1710553 and National Science Foundation ECCS-1711230.
E.A.O. and J.F. acknowledge ANID for the financial support (Becas de doctorado en el extranjero “Becas Chile”/2016 – No. 72170402 and 2015 – No. 72160487).

\appendix

\section{Entangled photon source}
\label{appendix:source}

\begin{figure*}[ht]
\includegraphics[width=0.9\textwidth]{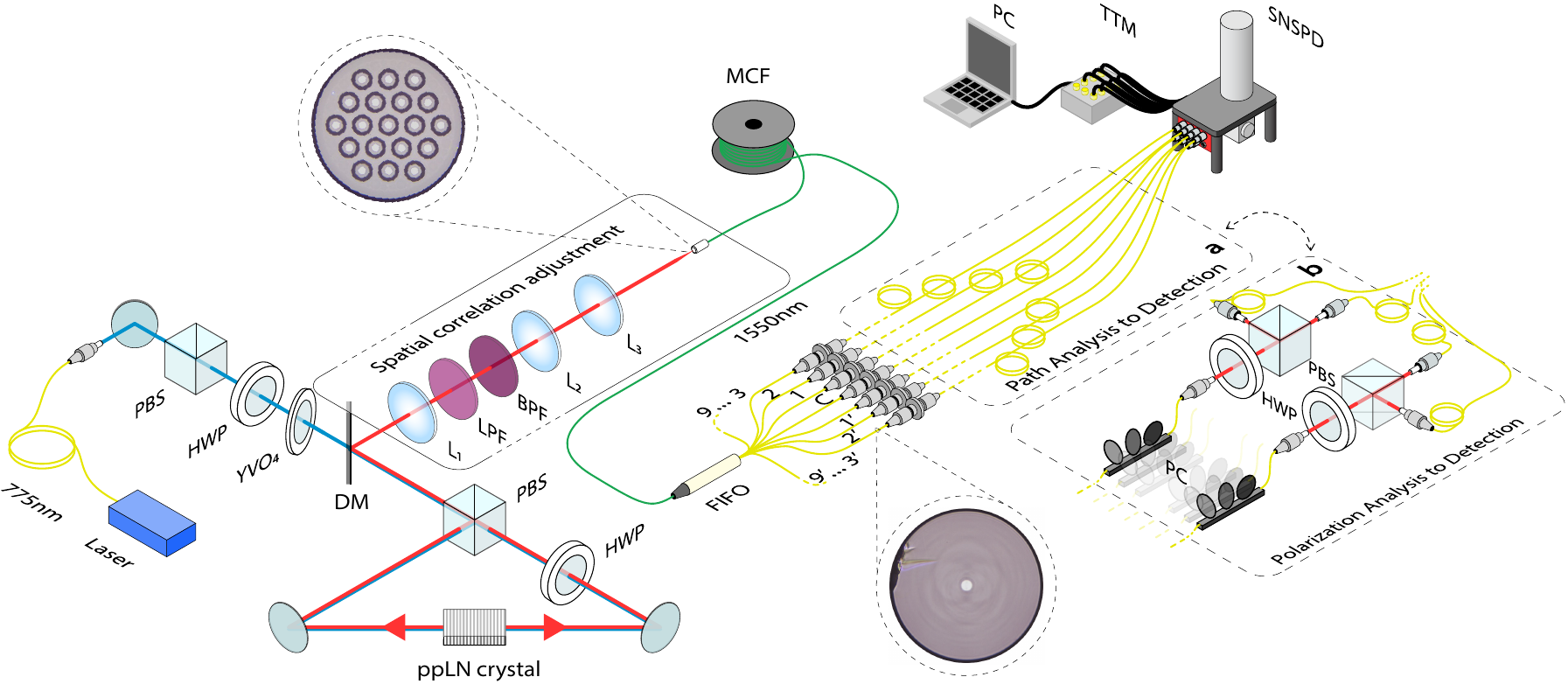}
\caption{Experimental setup.
A MgO:ppLN crystal is pumped in a Sagnac configuration by a continuous wave laser at $775.07$ nm (blue line) to create polarization entangled photon pairs with the center wavelength of $1550$ nm (red line).
The correlated photon pairs are distributed through opposite channels (cores) of the MCF with the detection modules for a) path visibility measurement and b) polarization visibility measurement.
The 19 cores are split into individual SMF via fan-in/fan-out (FIFO) device. 
In both scenarios the photons were detected using a superconducting nanowire single-photon detectors (SNSPD) with 4 channels.
The efficiency of the detectors depends on the photon polarization, so a polarization controller was used to optimize the detection efficiency before each channel. 
Abbreviations: PBS: polarizing beam-splitter, HWP: half-wave plate, YVO$_{4}$: yttrium orthovanadate plate, DM: dichroic mirror, LPF: long-pass filter, BPF: band-pass interference filter, PC: polarization controllers.
}
\label{fig:setup}
\end{figure*}

In our work, we used a polarization-entangled source based on type-0 phase matched SPDC process to distribute momentum correlation of photon pairs by a MCF.
The experimental set up is shown in Fig. \ref{fig:setup}.
The entanglement photon pairs are generated in a 40 mm long Magnesium Oxide doped periodically poled Lithium Niobate (MgO:ppLN) crystal with a poling period of $\Lambda = 19.2$ $\mu$m, pumped by a continuous-wave laser at the center wavelength $775.07$ nm.
The MgO:ppLN crystal was placed in a Sagnac interferometer without any additional active stabilization or compensation \cite{kim2006}.
If the pump photons propagate clockwise (counterclockwise) in the Sagnac loop, the down-converted photons will be horizontally polarized $\left\vert H_{s},H_{i}\right\rangle$ (vertically polarized $\left\vert V_{s},V_{i}\right\rangle$).
Under the perfect alignment, while the crystal is pumped in both directions, the down-converted photons will be in a superposition of horizontal and vertical polarization:
\begin{equation}
\left\vert \Phi^{+}\right\rangle = \frac{1}{\sqrt{2}}(\left\vert H_{s},H_{i}\right\rangle+\left\vert V_{\mathrm{s}},V_{i}\right\rangle).
\label{eq:state}
\end{equation}
The pump beam was spectrally cleaned with a long-pass filter ($\lambda_{cut-off} = 780$ nm).
An interference filter ($\lambda = 1550\pm3$ nm) was used to select the degenerated down-converted photons.

The far-field plane of the crystal was formed by the lens $\text L_{1}$ with focal length $f_{1} = 200\,$ mm.
Then, the photon pairs in the far-field plane were coupled into the MCF-cores by the imaging system consisting of the lenses $\text L_{2}$ ($f_{2} = 150$ mm) and $\text L_{3}$ ($f_{3} = 4.51$ mm).
This demagnified image of the down-converted photon pairs ensured that the emission cone-diameter covers all cores of a corresponding ring of cores.

By a scan of the crystal temperature, it can be verified that the emission cone, as seen in the Fourier plane, opens symmetrically around the center.
If the maximum count rate is reached at the same temperature for all the cores with the same center-to-core distances, then the end-face of the MCF is aligned with the emission cone.
Afterwards, the optimal temperature for each ring was chosen, at which the amount of coincidence counts between opposite cores is maximal.
In contrast, the amount of coincidences between non-opposite cores is low.
For the inner and the outer ring, the crystal temperature was set at $\text T_{1}=82.5^{\circ}$C and $\text T_{2}=82^{\circ}$C, respectively. 
For that temperature, the inner (outer) ring was completely illuminated by the SPDC emission cone.
Note how the outer ring has two core-to-core distances ($54.5$ $\mu$m and $64.5$ $\mu$m) making a circular projection more difficult. Nevertheless, our implementation showed reliable coupling of photons with highly preserved correlations even in this configuration.

\section{Allignment and characterization of the polarization state}
\label{appendix:allignment}

In order to quantify the maintaining of the polarization-entangled state $\left\vert \Phi^{+}\right\rangle$ between the spatial modes correlated in the far-field plane.
We selected two opposite cores and connected them to the polarization analyzing module.
The polarization visibility is measured in two mutually unbiased bases to quantify the polarization entanglement.
The polarization state of the photons was compensated in the H/V and in the D/A bases by fiber polarization controllers and by changing the polarization of the pump beam with a phase-plate YVO$_{4}$, respectively.
The visibility in HV-basis (the expression for DA-basis is analogous) is given by
\begin{equation}
    V_{HV} =\frac{C_{HH}+C_{VV}-C_{HV}-C_{VH}}{C_{HH}+C_{VV}+C_{HV}+C_{VH}}
\label{eq:polvis}
\end{equation}
where $C_{s_{1},s_{2}}$ is the number of coincidence counts for each polarization settings of the two opposite cores.
The same procedure is repeated for each pair of opposite cores.

\section{CHSH measurements}
\label{appendix:CHSH}

The CHSH inequality for two qubits state requires that\\$\mid S \mid\leq 2$.
Where the maximum value in quantum mechanics corresponds to the Tsirelson’s bound $2\sqrt{2}$.
Where the $S$ value is calculated by considering the coincidence counts of four combinations of settings in the polarization modules,
\begin{equation}
    S = E(a_{1},b_{1})+E(a_{2},b_{1})+E(a_{1},b_{2})-E(a_{2},b_{2})
\end{equation}
The correlation functions $E(a_{i},b_{j})$ are calculated from,
\begin{equation}
    E(a_{i},b_{j})=\frac{C(a_{i},b_{j})+C(a'_{i},b'_{j})-C(a'_{i},b_{j})-C(a_{i},b'_{j})}{C(a_{i},b_{j})+C(a'_{i},b'_{j})+C(a'_{i},b_{j})+C(a_{i},b'_{j})}
\nonumber
\end{equation}
$C(a_{i},b_{j})$ corresponds to the coincidence counts measure at the angles $(a_{i},b_{j})$. 
And we refer to the coincidence counts associated with the reflection output ports of the PBS to $a'_{i}$ and $b'_{j}$. 
We measured the S-value with polarization analyzers at the angles ($0^{\circ}$, $45^{\circ}$) and ($22.5^{\circ}$, $67.5^{\circ}$).

\section{Estimation of key rate}
\label{appendix:SKR}

The secret key rate for each pair of opposite cores given in the text has been estimated based on \cite{QKD_Ma,neumann2021}:
\begin{equation}
\label{eq:SKR}
\begin{split}
    R^{\mathrm{s}}_{m}& = \frac{1}{2}C_{HV}(1-(1+f)H_{2}(\frac{1-V_{HV}}{2}))\\
                 &\quad  +\frac{1}{2}C_{DA}(1-(1+f)H_{2}(\frac{1-V_{DA}}{2}))
\end{split}
\end{equation}
where $C_{HV}$ and $C_{DA}$ are the total number of coincidence counts in the HV and DA basis.
$H_{2}$ the binary shannon entropy and we used the bi-directional error correction $f=1.1$.

\nocite{*}

\bibliography{biblio}

\end{document}